\documentclass[a4paper,12pt]{article}

\usepackage{amsmath,amssymb}
\usepackage{bbm}
\usepackage{slashed}
\usepackage{feynmp}
\usepackage[nosort]{cite}
\usepackage{graphicx}
\usepackage[bookmarks=true]{hyperref}
\usepackage{axodraw}

\usepackage{ifpdf}
\ifpdf
\fi

\ifx\hypersetup\sadfkjashdfkxja\else
\hypersetup{pdftitle={}}
\hypersetup{pdfauthor={}}
\hypersetup{pdfsubject={}}
\hypersetup{pdfkeywords={}}
\hypersetup{plainpages=false}
\hypersetup{bookmarksnumbered=true}
\hypersetup{pdfstartview=FitH}
\hypersetup{pdfpagemode=UseNone}
\hypersetup{colorlinks=false}
\hypersetup{citebordercolor={.5 1 .5}}
\hypersetup{urlbordercolor={.5 1 1}}
\hypersetup{linkbordercolor={1 .7 .7}}
\fi

\newcommand{\Tr}{\mathop{\mathrm{Tr}}}

\def\be{\begin{equation}}
\def\ee{\end{equation}}
\def\bsp{\be\begin{split}}
\def\la{\langle}
\def\ra{\rangle}
\def\dag{\dagger}

\def\G{\Gamma}

\def\a{\alpha}
\def\b{\beta}
\def\g{\gamma}

\def\e{\epsilon}

\def\l{\lambda}

\def\o{\omega}

\renewcommand{\title}[1]{\vbox{\center\LARGE{#1}}\vspace{5mm}}
\renewcommand{\author}[1]{\vbox{\center\large{#1}}\vspace{5mm}}
\newcommand{\address}[1]{\vbox{\center\em#1}}
\newcommand{\email}[1]{\vbox{\center\tt#1}\vspace{5mm}}

\begin{document}
\begin{fmffile}{extremal1}
\bibliographystyle{utphys}

\begin{titlepage}
\hfill {\tt QMUL-PH-14-23}\\
\title{\vspace{1.0in} {\bf An Extremal Chiral Primary Three-Point
    Function at Two-loops in ABJ(M)}}
 
\author{Donovan Young}

\address{Centre for Research in String Theory\\
School of Physics and Astronomy\\
Queen Mary, University of London\\
Mile End Road, London E1 4NS, United Kingdom}

\email{d.young@qmul.ac.uk}

\abstract{I compute the leading correction to the structure constant
  for the three-point function of two length-two and one length-four
  chiral primary operators in planar ABJ(M) theory at weak 't Hooft
  coupling. The computation is reduced to four-loop propagator type
  Feynman integrals via a manifestly finite integration over the
  position of the length-four operator.}

\end{titlepage}

\section{Introduction and result}

This paper is a continuation of the works
\cite{Hirano:2012vz,Young:2013hda,Young:2014lka}, where it has been
pointed out that the ${\cal N}=6$ Chern-Simons matter theory of
Aharony, Bergman, Jafferis, and Maldacena (ABJM)
\cite{Aharony:2008ug,Aharony:2008gk} presents a nice lab for looking
at three-point functions because, unlike ${\cal N}=4$ SYM, the
protected chiral primary operators {\it don't} have protected
three-point functions. These operators are defined by
\be
{\cal O}^J_{A} =\frac{1}{\sqrt{J/2}}\left(\frac{k}{4\pi\sqrt{NM}}\right)^{J/2}
 ({\cal C}_{A})^{A_1\ldots A_{J/2}}_{B_1\ldots B_{J/2}} \,
\Tr \left(Y^{B_1} Y^\dag_{A_1} \cdots Y^{B_{J/2}} Y^\dag_{A_{J/2}}\right),
\ee
where the $Y^{A_i=1,2,3,4}$ are the bifundamental $(N,\bar M)$ complex
scalars of the theory, $k$ is the Chern-Simons level and ${\cal C}_A$
is a traceless symmetric tensor. Their conformally-fixed three-point
functions
\be
\bigl\la {\cal O}_1(x_1) {\cal O}_2(x_2) {\cal O}_3(x_3) \bigr\ra = 
\frac{1}{(4\pi)^\g}
\frac{C_{123}(\l,\hat\l)}
{|x_{12}|^{\g_3} |x_{23}|^{\g_1} |x_{31}|^{\g_2}},
\ee
where $\gamma_i = (\sum_jJ_j-2J_i)/2$, $\g=\g_1+\g_2+\g_3$, and
$x_{ij}=x_i-x_j$, include non-trivial dependence of the structure
constant $C_{123}$ on the 't Hooft couplings $\l = N/k$,
$\hat\l=M/k$. At strong coupling, and in the ABJM case where $N=M$,
supergravity tells us that \cite{Bastianelli:1999en,Hirano:2012vz}
\bsp
&C_{123}(\l\gg 1) =\\
&\frac{1}{N} \left(\frac{\l}{2\pi^2}\right)^{1/4}
\frac{\prod_{i=1}^3 
\sqrt{J_i+1}\,(J_i/2)!\,\G(\g_i/2+1) }{\Gamma(\g/2+1)} \\
&\sum_{p=0}^{\g_3}
\frac{\left({\cal C}_1\right)^{I_1\ldots I_p I_{p+1}\ldots I_{J_1/2}}
_{K_1 \ldots K_{\g_3 -p} K_{\g_3-p+1} \ldots K_{J_1/2}}
\left({\cal C}_2\right)^{K_1 \ldots K_{\g_3 -p} L_1\ldots
L_{\g_1-J_2/2+p}}_{I_1\ldots I_p M_1 \ldots M_{J_2/2-p}}
\left({\cal C}_3\right)^{K_{\g_3-p+1} \ldots K_{J_1/2}M_1 \ldots
  M_{J_2/2-p}}
_{I_{p+1}\ldots I_{J_1/2}  L_1\ldots L_{\g_1-J_2/2+p}}}
{p! (\g_3-p)! (\g_1-J_2/2+p)!
(J_2/2-p)!(\g_2-J_1/2+p)! (J_1/2-p)!},
\end{split}
\ee 
and suggests a series of interpolating functions corresponding to the
independent ways the ${\cal C}$-tensors can be contracted. 

The case of extremal correlators -- when $J_3 = J_1+J_2$ -- offers a
dramatic simplification. In this case we have a single possible
contraction (labelled ``tree'' for reasons to be explained directly)
and obtain
\be\label{exstr}
C_{123}^{\text{extremal}}(\l\gg 1)=
\frac{1}{N} \left(\frac{\l}{2\pi^2}\right)^{1/4}
\sqrt{(J_1+1)(J_2+1)(J_3+1)}\, \la {\cal C}_{1}{\cal
  C}_{2}{\cal C}_{3}\ra_{\text{tree}}.
\ee
This expression is similar to the tree-level expression
\be
C_{123}(\l \to 0) = C_F
\sqrt{(J_1/2)(J_2/2)(J_3/2)}\, \la {\cal C}_1\,{\cal
  C}_2\,{\cal C}_3\ra_{\text{tree}}, 
\ee
and thus suggests a simplification compared to the non-extremal
case. In this paper I compute the structure constant at leading loop
order for the simplest such extremal correlator consisting of one
length-four operator and two length-two operators. The result
is\footnote{We choose to factor out the colour factor
  $C_F = 1/N+1/M$ and a factor of $1/\sqrt{2}$ so that $C_{123}=1$ at
  tree-level.}
\be\label{mainresult}
C_{123}\bigl|_{{\cal O}(\l^2)} = \frac{\pi^2}{6}\left(\l^2+\hat\l^2\right). 
\ee
For the ABJM case where $N=M$ the strong coupling result is (\ref{exstr})
\be
C_{123}(\l\to\infty) =
\left(\frac{\l}{2\pi^2}\right)^{1/4} \frac{3\sqrt{5}}{2\sqrt{2}},
\ee
and thus $C_{123}$ interpolates between $1+\l^2\pi^2/3$ and this
value. It remains a very interesting direction of future research to
determine this interpolating function exactly.

In the following sections of the paper I review the method of
calculation first presented in \cite{Young:2014lka}, and give an
account of the various Feynman diagrams contributing to the
result. The appendices contain the details of the reduction to master
integrals and the presentation of the master integrals themselves. We
have used the dimensional regularization scheme adopted in
\cite{Chen:1992ee,Minahan:2009wg} where $d=2\o=3-2\e$ and all
numerators are reduced to scalar products at the physical dimension
first, while loop integration proceeds as usual using dimensional
regularization. The conventions used here are identical to those used
in very similar contexts in
\cite{Minahan:2009wg,Young:2013hda,Young:2014lka} and the reader is
directed to these works for statements of the action, Feynman rules
and other details.

\section{Spacetime point integration method}

In \cite{Young:2014lka} a novel method for computing three-point
functions of protected operators was suggested. The idea is to exploit
the finiteness of the three-point function by integrating over one of
the spacetime points where one of the three operators is sitting,
using the same dimensional regularization scheme used in the loop
computation. This reduces the required Feynman integrals to
propagator-type, which are more easily evaluated. 

In the case of two length-two and one length-four operator in ABJM,
this integration is itself convergent if we choose the point where the
length-four operator sits to integrate over. We will look at the
following specific operators
\bsp
&{\cal O}_1 = \frac{k}{4\pi\sqrt{MN}}\Tr( Y^1Y^\dag_2),\\
&{\cal O}_2 = \frac{k^2}{\sqrt{2}(4\pi)^2 MN}
\Bigl(\Tr( Y^2Y^\dag_1Y^4Y^\dag_3 )
+ \Tr( Y^\dag_1Y^2Y^\dag_3Y^4 ) \Bigr),\\
&{\cal O}_3 = \frac{k}{4\pi\sqrt{MN}}\Tr( Y^3Y^\dag_4).
\end{split}
\ee
The three-point function is
\be
\bigl\la {\cal O}_1(0) {\cal O}_2(x) {\cal O}_3(y) \bigr\ra 
= \frac{1}{\sqrt{2}}\left(\frac{1}{N}+\frac{1}{M}\right)
\frac{1}{(4\pi)^4} \frac{\hat C_{123}(\l,\hat\l)}{x^2(x-y)^2}.
\ee
Integrating over $x$, we obtain
\be
\int d^3 x \bigl\la {\cal O}_1(0) {\cal O}_2(x) {\cal O}_3(y) \bigr\ra
= \frac{1}{\sqrt{2}}\left(\frac{1}{N}+\frac{1}{M}\right)
\frac{\hat C_{123}(\l,\hat\l)}{256\pi |y|},
\ee
and so 
\be\label{unrenorm}
\hat C_{123}(\l,\hat\l) =
256\pi|y| \sqrt{2}\left(\frac{1}{N}+\frac{1}{M}\right)^{-1} 
\int d^3 x \bigl\la {\cal O}_1(0) {\cal O}_2(x) {\cal O}_3(y) \bigr\ra.
\ee
Thus the task is to calculate $\int d^3 x \bigl\la {\cal O}_1(0) {\cal
  O}_2(x) {\cal O}_3(y) \bigr\ra$, which in momentum space amounts to
setting to zero the momentum flowing into the central operator. We
thus obtain four-loop propagator-type diagrams which we tackle in the
following section.

The structure constant is renormalized by the two-point functions of
the three operators according to
\be\label{renorm}
C_{123}\bigl|_{{\cal O}(\l^2)} = \left[\hat C_{123}
-\frac{1}{2}\sum_{i=1}^3 g_i
\right]_{{\cal O}(\l^2)}.
\ee
where
\bsp\label{twoptnorm}
&\bigl\la {\cal O}_1 (x) {\cal O}_1(0) \bigr\ra=
\frac{g_1(\l,\hat\l)}{(4\pi |x|)^2},\quad
\bigl\la {\cal O}_2 (x) {\cal O}_2(0) \bigr\ra=
\frac{g_2(\l,\hat\l)}{(4\pi |x|)^4},\\
&\bigl\la {\cal O}_3 (x) {\cal O}_3(0) \bigr\ra=
\frac{g_3(\l,\hat\l)}{(4\pi |x|)^2},
\end{split}
\ee
and where $g_i(0,0) =1$.

\section{Feynman diagrammatics}

We will require the two-loop decorations of the (integrated)
tree-level correlator
\begin{center}
\begin{fmfgraph*}(70,70)
\fmfleft{v1}
\fmfright{v2}
\fmf{plain,right=1}{v1,va,v2,va,v1}
\fmfv{decor.shape=circle,decor.filled=30,decor.size=9}{v1}
\fmfv{decor.shape=circle,decor.filled=30,decor.size=9}{v2}
\end{fmfgraph*}
\end{center}
where we have indicated the length-two operators with grey blobs. The
first thing to notice is that diagrams in which only one of the two
loops are decorated
\vspace{-0.4cm}
\begin{center}
\begin{fmfgraph*}(90,90)
\fmfleft{v1}
\fmfright{v2}
\fmf{plain,right=0.8}{v1,va,vb,v2,vb,va,v1}
\fmfv{decor.shape=circle,decor.filled=30,decor.size=9}{v1}
\fmfv{decor.shape=circle,decor.filled=30,decor.size=9}{v2}
\fmfv{d.sh=circle,l.d=0, d.f=10,d.si=.3w,l=$2$}{va}
\end{fmfgraph*}
\end{center}
\vspace{-0.4cm}
are removed by the $g_1$ and $g_3$ terms of the renormalization
(\ref{renorm}). We thus move on to the diagrams in which decorations
connect the right and left loops. Many potential diagrams evaluate to
zero via two mechanisms. The first is that a gauge field line drawn
across two legs of an operator (from one scalar-scalar-gauge vertex
on each leg) will produce zero as long as it can be contracted to the
operator site (i.e. without encountering other vertices along the way)
\cite{Young:2013hda}. Examples of these types of diagrams
are\footnote{Note that gauge fields are represented via wiggly lines,
  fermions via dashes and scalar fields via plain lines. The Feynman
  rules, action, and conventions used here have been detailed
  elsewhere \cite{Young:2013hda,Minahan:2009wg}. }
\begin{center}
\vspace{-1.5cm}
\[
\parbox{20mm}{\vspace{1.5cm}
\begin{fmfgraph*}(40,70)
\fmftop{v1}
\fmfleft{v2}
\fmfright{v3}
\fmfv{decoration.shape=circle,decoration.filled=30,decor.size=9}{v1}
\fmf{plain}{v1,vc1}
\fmf{plain}{vc1,vc3}
\fmf{plain}{vc3,v2}
\fmf{plain}{v1,vc2}
\fmf{plain}{vc2,vc4}
\fmf{plain}{vc4,v3}
\fmffreeze
\fmf{photon,right=0.15}{vc1,vc2}
\fmffreeze
\fmfposition
\end{fmfgraph*}}
=
\hspace{0.5cm}
\parbox{20mm}{\vspace{1.5cm}
\begin{fmfgraph*}(40,70)
\fmftop{v1}
\fmfleft{v2}
\fmfright{v3}
\fmfv{decoration.shape=circle,decoration.filled=30,decor.size=9}{v1}
\fmf{plain}{v1,vc1}
\fmf{plain}{vc1,vc3}
\fmf{plain}{vc3,v2}
\fmf{plain}{v1,vc2}
\fmf{plain}{vc2,vc4}
\fmf{plain}{vc4,v3}
\fmffreeze
\fmf{photon,right=0.15}{vc1,vc2}
\fmf{photon,right=0.15}{vc3,vc4}
\fmffreeze
\fmfposition
\end{fmfgraph*}}
=
\hspace{0.5cm}
\parbox{20mm}{\vspace{.4cm}
\begin{fmfgraph*}(40,40)
\fmftop{v1}
\fmfstraight
\fmfbottom{v2,v3,v4}
\fmfv{decoration.shape=circle,decoration.filled=30,decor.size=9}{v1}
\fmf{plain}{v1,vc1}
\fmf{plain}{vc1,v2}
\fmf{plain}{v1,vc2}
\fmf{plain}{vc2,vcc2}
\fmf{plain}{vcc2,v3}
\fmf{plain}{v1,vc3}
\fmf{plain}{vc3,v4}
\fmffreeze
\fmf{photon}{vc1,vc2}
\fmf{photon}{vcc2,vc3}
\end{fmfgraph*}}=~~0.\]
\vspace{-1cm}
\end{center}
The second mechanism is encountered when a
single gauge field connects to the left or right loop
\begin{center}
\[
\parbox{20mm}{
\begin{fmfgraph*}(70,70)
\fmfleft{v1}
\fmfright{v2}
\fmfbottom{b1}
\fmf{plain,right=0.2}{v1,va}
\fmf{plain,right=0.2}{va,v2}
\fmf{plain,right=0.3}{v2,v1}
\fmf{wiggly}{b1,va}
\fmfv{decor.shape=circle,decor.filled=30,decor.size=9}{v2}
\end{fmfgraph*}}~\qquad +~~~
\parbox{20mm}{
\begin{fmfgraph*}(70,70)
\fmfleft{v1}
\fmfright{v2}
\fmftop{t1}
\fmf{plain,left=0.2}{v1,va}
\fmf{plain,left=0.2}{va,v2}
\fmf{plain,left=0.3}{v2,v1}
\fmf{wiggly}{t1,va}
\fmfv{decor.shape=circle,decor.filled=30,decor.size=9}{v2}
\end{fmfgraph*}}~\qquad=~0,
\]
\vspace{0.05cm}
\end{center}
here the sum of connecting the lone gauge field to the top line of the
loop and to the bottom produces zero via a simple integration by
parts. Note that in the above we have not shown the left loop where
the gauge field originates. We are then left with a series of
non-vanishing diagrams, which have been collected in table
\ref{tab:maindiags}. 

\begin{table}
\begin{tabular}{ccccc}
\parbox{20mm}{
\begin{fmfgraph*}(50,50)
\fmfleft{v1}
\fmfright{v5}
\fmftop{t2,t4,t3}
\fmfbottom{b1}
\fmf{plain,left=0.5, tension=0.75}{v1,v2}
\fmf{plain,left=0.4, tension=1}{v2,v3}
\fmf{plain,left=0.4, tension=1}{v3,v4}
\fmf{plain,left=0.5, tension=0.75}{v4,v5}
\fmf{plain,left=1., tension=1}{v5,v3,v1}
\fmf{phantom,tension=2.8}{t2,v2}
\fmf{phantom,tension=2.8}{t3,v4}
\fmf{phantom,tension=1.05}{v3,b1}
\fmf{wiggly,left=.3}{v2,v4}
\fmf{wiggly,left=.9}{v2,v4}
\fmfv{decor.shape=circle,decor.filled=30,decor.size=9}{v1}
\fmfv{decor.shape=circle,decor.filled=30,decor.size=9}{v5}
\end{fmfgraph*}}&
\hspace{0.5cm}
\parbox{20mm}{
\begin{fmfgraph*}(50,50)
\fmfleft{v1}
\fmfright{v5}
\fmftop{t2,t4,t3}
\fmfbottom{b2,b1,b3}
\fmf{plain,left=0.35, tension=0.75}{v1,vb}
\fmf{plain,left=0.2, tension=114.25}{vb,v2}
\fmf{plain,left=0.4, tension=1}{v2,v3}
\fmf{plain,left=1., tension=1}{v3,v5}
\fmf{plain,left=0.35, tension=.75}{v5,vc}
\fmf{plain,left=0.2, tension=114.25}{vc,v4}
\fmf{plain,left=0.4, tension=1}{v4,v3}
\fmf{plain,left=1., tension=1}{v3,v1}
\fmf{phantom,tension=3.}{t4,v2}
\fmf{phantom,tension=6}{t2,vb}
\fmf{phantom,tension=3.}{b1,v4}
\fmf{phantom,tension=6}{b3,vc}
\fmf{wiggly,right=2.3,tension=1}{vc,v2}
\fmf{wiggly,left=2.3,tension=1}{v4,vb}
\fmfv{decor.shape=circle,decor.filled=30,decor.size=9}{v1}
\fmfv{decor.shape=circle,decor.filled=30,decor.size=9}{v5}
\end{fmfgraph*}}&
\hspace{0.5cm}
\parbox{20mm}{
\begin{fmfgraph*}(50,50)
\fmfleft{v1}
\fmfright{v5}
\fmftop{t2,t4,t3}
\fmfbottom{b1}
\fmf{plain,left=0.35, tension=0.75}{v1,vb}
\fmf{plain,left=0.1, tension=0.25}{vb,v2}
\fmf{plain,left=0.4, tension=1}{v2,v3}
\fmf{plain,left=0.4, tension=1}{v3,v4}
\fmf{plain,left=0.5, tension=0.75}{v4,v5}
\fmf{plain,left=1., tension=1}{v5,v3,v1}
\fmf{phantom,tension=2.8}{t2,v2}
\fmf{phantom,tension=1.}{t2,vb}
\fmf{phantom,tension=2.8}{t3,v4}
\fmf{phantom,tension=1.05}{v3,b1}
\fmf{wiggly,left=.3}{v2,v4}
\fmf{wiggly,left=.7,tension=0.1}{vb,v4}
\fmfv{decor.shape=circle,decor.filled=30,decor.size=9}{v1}
\fmfv{decor.shape=circle,decor.filled=30,decor.size=9}{v5}
\end{fmfgraph*}}&
\hspace{0.5cm}
\parbox{20mm}{
\begin{fmfgraph*}(50,50)
\fmfleft{v1}
\fmfright{v5}
\fmftop{t2,t4,t3}
\fmfbottom{b2,b1,b3}
\fmf{plain,left=0.35, tension=0.75}{v1,vb}
\fmf{plain,left=0.2, tension=114.25}{vb,v2}
\fmf{plain,left=0.4, tension=1}{v2,v3}
\fmf{plain,left=1., tension=1}{v3,v5}
\fmf{plain,left=0.35, tension=.75}{v5,vc}
\fmf{plain,left=0.2, tension=4.25}{vc,v4}
\fmf{plain,left=0.3, tension=1}{v4,v3}
\fmf{plain,left=1., tension=1}{v3,v1}
\fmf{phantom,tension=3.}{t4,v2}
\fmf{phantom,tension=6}{t2,vb}
\fmf{phantom,tension=3.}{b1,v4}
\fmf{phantom,tension=6}{b3,vc}
\fmf{wiggly,right=1.93,tension=1}{vc,v2}
\fmf{wiggly,left=1.93,tension=1}{v4,vb}
\fmfv{decor.shape=circle,decor.filled=30,decor.size=9}{v1}
\fmfv{decor.shape=circle,decor.filled=30,decor.size=9}{v5}
\end{fmfgraph*}}&
\hspace{0.75cm}
\parbox{20mm}{
\begin{fmfgraph*}(50,50)
\fmfleft{v1}
\fmfright{v5}
\fmftop{t2,t4,t3}
\fmfbottom{b2,b1,b3}
\fmf{plain,left=0.5, tension=0.75}{v1,v2}
\fmf{plain,left=0.25, tension=2}{v2,v3}
\fmf{plain,left=0.4, tension=1}{v3,v4}
\fmf{plain,left=0.4, tension=0.75}{v4,v5}
\fmf{plain,left=1.1, tension=1}{v5,v3}
\fmf{plain,left=0.6, tension=0.75}{v3,v6}
\fmf{plain,left=0.4, tension=1}{v6,v1}
\fmf{phantom,tension=2.2}{t2,v2}
\fmf{phantom,tension=4.2}{t3,v4}
\fmf{phantom,tension=.35}{v3,b1}
\fmf{phantom,tension=6.8}{v6,b2}
\fmf{wiggly,left=.8}{v2,v4}
\fmf{wiggly,left=2.05}{v6,v4}
\fmfv{decor.shape=circle,decor.filled=30,decor.size=9}{v1}
\fmfv{decor.shape=circle,decor.filled=30,decor.size=9}{v5}
\end{fmfgraph*}}\\
\parbox{20mm}{\vspace{1cm}
\begin{fmfgraph*}(50,50)
\fmfleft{v1}
\fmfright{v5}
\fmftop{t2,t4,t3}
\fmfbottom{b2,b1,b3}
\fmf{plain,left=0.35, tension=0.75}{v1,vb}
\fmf{plain,left=0.2, tension=4.25}{vb,v2}
\fmf{plain,left=0.3, tension=1}{v2,v3}
\fmf{plain,left=1., tension=1}{v3,v5}
\fmf{plain,left=0.35, tension=.75}{v5,vc}
\fmf{plain,left=0.2, tension=4.25}{vc,v4}
\fmf{plain,left=0.3, tension=1}{v4,v3}
\fmf{plain,left=1., tension=1}{v3,v1}
\fmf{phantom,tension=3.}{t4,v2}
\fmf{phantom,tension=6}{t2,vb}
\fmf{phantom,tension=3.}{b1,v4}
\fmf{phantom,tension=6}{b3,vc}
\fmf{wiggly,right=1.93,tension=1}{vc,v2}
\fmf{wiggly,left=1.93,tension=1}{v4,vb}
\fmfv{decor.shape=circle,decor.filled=30,decor.size=9}{v1}
\fmfv{decor.shape=circle,decor.filled=30,decor.size=9}{v5}
\end{fmfgraph*}}&
\hspace{0.5cm}
\parbox{20mm}{\vspace{1cm}
\begin{fmfgraph*}(50,50)
\fmfleft{v1}
\fmfright{v5}
\fmftop{t2,t4,t3}
\fmfbottom{b2,b1,b3}
\fmf{plain,left=0.35, tension=0.75}{v1,vb}
\fmf{plain,left=0.2, tension=114.25}{vb,v2}
\fmf{plain,left=0.4, tension=1}{v2,v3}
\fmf{plain,left=1., tension=1}{v3,v5}
\fmf{plain,left=0.35, tension=.75}{v5,vc}
\fmf{plain,left=0.2, tension=114.25}{vc,v4}
\fmf{plain,left=0.4, tension=1}{v4,v3}
\fmf{plain,left=1., tension=1}{v3,v1}
\fmf{phantom,tension=3.}{t4,v2}
\fmf{phantom,tension=6}{t2,vb}
\fmf{phantom,tension=3.}{b1,v4}
\fmf{phantom,tension=6}{b3,vc}
\fmf{dashes,right=2.3,tension=1}{vc,v2}
\fmf{dashes,left=2.3,tension=1}{v4,vb}
\fmfv{decor.shape=circle,decor.filled=30,decor.size=9}{v1}
\fmfv{decor.shape=circle,decor.filled=30,decor.size=9}{v5}
\end{fmfgraph*}}&
\hspace{0.5cm}
\parbox{20mm}{\vspace{1cm}
\begin{fmfgraph*}(50,50)
\fmfleft{v1}
\fmfright{v3}
\fmf{plain,left=0.3, tension=1}{v1,v4,v3,v2,v1}
\fmffixed{(0,33)}{v2,v4}
\fmf{plain,left=.93}{v2,v5}
\fmf{dashes,left=.93}{v5,v4}
\fmf{plain,right=.93}{v2,v5}
\fmf{dashes,right=.93}{v5,v4}
\fmfv{decor.shape=circle,decor.filled=30,decor.size=9}{v1}
\fmfv{decor.shape=circle,decor.filled=30,decor.size=9}{v3}
\end{fmfgraph*}}&
\hspace{0.5cm}
\parbox{20mm}{\vspace{1cm}
\begin{fmfgraph*}(50,50)
\fmfleft{v1}
\fmfright{v3}
\fmftop{t1,t2}
\fmfbottom{b1,b2}
\fmf{plain,left=.8, tension=1}{v1,v4,v5,v3,v4,v1}
\fmf{plain}{v4,v5}
\fmf{plain,right=.93}{v4,v5}
\fmf{phantom,tension=1}{t1,v4}
\fmf{phantom,tension=4}{t2,v5}
\fmfv{decor.shape=circle,decor.filled=30,decor.size=9}{v1}
\fmfv{decor.shape=circle,decor.filled=30,decor.size=9}{v3}
\end{fmfgraph*}}
\end{tabular}\vspace{0.5cm}
\caption{Decorations of the three-point correlator. Unique diagrams
  obtained through reflection about the horizontal and vertical axes
  must also be considered.}
\label{tab:maindiags}
\end{table}

\begin{table}
\begin{tabular}{ccccc}
\parbox{20mm}{
\begin{fmfgraph*}(50,50)
\fmfleft{v1}
\fmfright{v3}
\fmf{plain,left=0.3, tension=1}{v1,v4,v3,v2,v1}
\fmf{plain,left=1., tension=1}{v1,v3,v1}
\fmffixed{(0,33)}{v2,v4}
\fmf{wiggly,left=.3}{v2,v4,v2}
\fmfv{decor.shape=circle,decor.filled=30,decor.size=9}{v1}
\fmfv{decor.shape=circle,decor.filled=30,decor.size=9}{v3}
\end{fmfgraph*}}&
\hspace{0.5cm}
\parbox{20mm}{
\begin{fmfgraph*}(50,50)
\fmfleft{v1}
\fmftop{t}
\fmfright{v2}
\fmfbottom{b1,b2}
\fmf{phantom,tension=2.95}{t,va}
\fmf{phantom,tension=2.15}{b1,vb1}
\fmf{phantom,tension=2.15}{b2,vb2}
\fmf{plain,left=0.3}{v1,va}
\fmf{plain,left=0.3}{va,v2}
\fmf{plain,right=0.25}{v1,vb1}
\fmf{plain,right=0.25}{vb1,vb2}
\fmf{plain,right=0.25}{vb2,v2}
\fmffreeze
\fmf{wiggly}{va,vb1}
\fmf{wiggly}{va,vb2}
\fmf{plain,left=1., tension=1}{v1,v2}
\fmf{plain,left=1.2, tension=1}{v2,v1}
\fmfv{decor.shape=circle,decor.filled=30,decor.size=9}{v1}
\fmfv{decor.shape=circle,decor.filled=30,decor.size=9}{v2}
\end{fmfgraph*}}&
\hspace{0.5cm}
\parbox{20mm}{
\begin{fmfgraph*}(50,50)
\fmfleft{v1}
\fmftop{t1,t2,t3}
\fmfright{v2}
\fmfbottom{b}
\fmf{phantom,tension=1.0}{t1,va1}
\fmf{phantom,tension=1.1}{t2,va2}
\fmf{phantom,tension=1.0}{t3,va3}
\fmf{phantom,tension=2.95}{b,vb}
\fmf{plain,left=0.17}{v1,va1}
\fmf{plain,left=0.15}{va1,va2}
\fmf{plain,left=0.15}{va2,va3}
\fmf{plain,left=0.17}{va3,v2}
\fmf{plain,right=0.3}{v1,vb}
\fmf{plain,right=0.3}{vb,v2}
\fmffreeze
\fmf{wiggly,left=0.75}{va1,va3}
\fmf{wiggly}{va2,vb}
\fmf{plain,left=1.3, tension=1}{v1,v2}
\fmf{plain,left=1., tension=1}{v2,v1}
\fmfv{decor.shape=circle,decor.filled=30,decor.size=9}{v1}
\fmfv{decor.shape=circle,decor.filled=30,decor.size=9}{v2}
\end{fmfgraph*}}&
\hspace{0.5cm}
\parbox{20mm}{
\begin{fmfgraph*}(50,50)
\fmfleft{v1}
\fmftop{t1,t2,t3}
\fmfright{v2}
\fmfbottom{b}
\fmf{phantom,tension=1.0}{t1,va1}
\fmf{phantom,tension=1.1}{t2,va2}
\fmf{phantom,tension=1.0}{t3,va3}
\fmf{phantom,tension=2.95}{b,vb}
\fmf{plain,left=0.17}{v1,va1}
\fmf{plain,left=0.15}{va1,va2}
\fmf{plain,left=0.15}{va2,va3}
\fmf{plain,left=0.17}{va3,v2}
\fmf{plain,right=0.3}{v1,vb}
\fmf{plain,right=0.3}{vb,v2}
\fmffreeze
\fmf{wiggly}{va1,vcen}
\fmf{wiggly}{va3,vcen}
\fmf{wiggly}{vb,vcen}
\fmf{plain,left=1.2, tension=1}{v1,v2}
\fmf{plain,left=1., tension=1}{v2,v1}
\fmfv{decor.shape=circle,decor.filled=30,decor.size=9}{v1}
\fmfv{decor.shape=circle,decor.filled=30,decor.size=9}{v2}
\end{fmfgraph*}}&
\hspace{0.5cm}
\parbox{20mm}{
\begin{fmfgraph*}(50,50)
\fmfleft{v1}
\fmfright{v3}
\fmf{plain,left=0.3, tension=1}{v1,v4,v3,v2,v1}
\fmffixed{(0,33)}{v2,v4}
\fmf{wiggly}{v2,vc,v4}
\fmf{plain,left=1., tension=1}{v1,v3,v1}
\fmfv{d.sh=circle,l.d=0, d.f=empty,d.si=.25w,l=$1$}{vc}
\fmfv{decor.shape=circle,decor.filled=30,decor.size=9}{v1}
\fmfv{decor.shape=circle,decor.filled=30,decor.size=9}{v3}
\end{fmfgraph*}}\\
\parbox{20mm}{\vspace{1cm}
\begin{fmfgraph*}(50,50)
\fmfleft{v1}
\fmfright{v3}
\fmf{plain,left=1., tension=1}{v1,va}
\fmf{dashes,left=1., tension=1}{va,vb}
\fmf{plain,left=1., tension=1}{vb,v3}
\fmf{plain,right=1., tension=1}{v1,va}
\fmf{dashes,right=1., tension=1}{va,vb}
\fmf{plain,right=1., tension=1}{vb,v3}
\fmf{plain,left=.75, tension=1}{v1,v3,v1}
\fmfv{decor.shape=circle,decor.filled=30,decor.size=9}{v1}
\fmfv{decor.shape=circle,decor.filled=30,decor.size=9}{v3}
\end{fmfgraph*}}&
\hspace{0.5cm}
\parbox{20mm}{\vspace{1cm}
\begin{fmfgraph*}(50,50)
\fmfleft{v1}
\fmfright{v3}
\fmf{plain,left=0.3, tension=1}{v1,v4,v3,v2,v1}
\fmffixed{(0,30)}{v2,v4}
\fmf{plain,left=1., tension=1}{v1,v3,v1}
\fmfv{d.sh=circle,l.d=0, d.f=empty,d.si=.25w,l=$2$}{v4}
\fmfv{decor.shape=circle,decor.filled=30,decor.size=9}{v1}
\fmfv{decor.shape=circle,decor.filled=30,decor.size=9}{v3}
\end{fmfgraph*}}&
\hspace{0.5cm}
\parbox{20mm}{\vspace{1cm}
\begin{fmfgraph*}(50,50)
\fmfleft{v1}
\fmfright{v3}
\fmf{plain,left=0.93, tension=1}{v1,v4}
\fmf{plain, tension=1}{v1,v4}
\fmf{plain,right=0.93, tension=1}{v1,v4}
\fmf{plain,left=0.93, tension=1}{v4,v3}
\fmf{plain, tension=1}{v4,v3}
\fmf{plain,right=0.93, tension=1}{v4,v3}
\fmf{plain,left=1., tension=1}{v3,v1}
\fmfv{decor.shape=circle,decor.filled=30,decor.size=9}{v1}
\fmfv{decor.shape=circle,decor.filled=30,decor.size=9}{v3}
\end{fmfgraph*}}&
\hspace{0.5cm}
\parbox{20mm}{\vspace{1cm}
\begin{fmfgraph*}(50,50)
\fmfleft{v1}
\fmfright{v3}
\fmftop{t1}
\fmfbottom{b1}
\fmf{plain,left=0.3, tension=1}{v1,v4,v3}
\fmf{plain, tension=1}{v1,v2,v3}
\fmf{plain,right=.3, tension=1}{v1,v5,v3}
\fmf{plain,right=1, tension=1}{v1,v3}
\fmf{phantom,tension=5}{b1,v5}
\fmf{phantom,tension=5}{t1,v4}
\fmf{wiggly}{v2,v5}
\fmf{wiggly}{v2,v4}
\fmfv{decor.shape=circle,decor.filled=30,decor.size=9}{v1}
\fmfv{decor.shape=circle,decor.filled=30,decor.size=9}{v3}
\end{fmfgraph*}}&
\end{tabular}
\vspace{0.5cm}
\caption{The Feynman diagrams which contribute to the two-point
  function of the length-four operator. The various decorations must
  be considered on all sets of adjacent legs. Unique diagrams obtained
  through reflection about the vertical axis must also be
  considered.}
\label{tab:twoptdiags}
\end{table}
These diagrams are evaluated using the Feynman rules published in
\cite{Minahan:2009wg} (note that these Feynman rules capture only the
planar contributions which is sufficient for the present calculation)
and used already in \cite{Young:2013hda,Young:2014lka}. The resulting
integrals are then reduced to master integrals using the LiteRed
package of Roman N. Lee \cite{Lee:2013mka}, and in particular the
``p4.zip'' basis provided on his
website\footnote{\url{http://www.inp.nsk.su/~lee/programs/LiteRed/}}. One
then requires the $\e$-expansion of these master integrals generically
out to ${\cal O}(\e^2)$. In four-dimensions these integrals have been
evaluated to high-orders in the $\e$-expansion \cite{Lee:2011jt},
starting with the original work \cite{Baikov:2010hf} which used the
glue-and-cut symmetry, and using the method of dimensional-recurrence
and analyticity also pioneered by Lee \cite{Lee:2009dh}. The author is
very grateful to Prof. Lee for providing the three-dimensional epsilon
expansion of the required integrals out to 500-digit accuracy which
have allowed this computation to be executed \cite{LeePriv}.

We must also evaluate the finite renormalization of the two-point
function for the length-four operator. The diagrams required are shown
in table \ref{tab:twoptdiags}. They are mostly captured by the
three-loop propagator diagrams encountered in \cite{Young:2014lka} but
the final two diagrams require the same four-loop integrals discussed
above. 

We conclude this section with a statement of the results. The
reduction of the individual diagrams to master integrals is given in
appendix \ref{app:red}, and in appendix \ref{app:4lmasters} we
quote the results for the master integrals in an $\e$-expansion about
three-dimensions. In table \ref{tab:maindiags} the following diagrams
add to zero at ${\cal O}(\e^0)$
\vspace{0.4cm}
\begin{center}
\[
\parbox{20mm}{
\begin{fmfgraph*}(50,50)
\fmfleft{v1}
\fmfright{v5}
\fmftop{t2,t4,t3}
\fmfbottom{b1}
\fmf{plain,left=0.5, tension=0.75}{v1,v2}
\fmf{plain,left=0.4, tension=1}{v2,v3}
\fmf{plain,left=0.4, tension=1}{v3,v4}
\fmf{plain,left=0.5, tension=0.75}{v4,v5}
\fmf{plain,left=1., tension=1}{v5,v3,v1}
\fmf{phantom,tension=2.8}{t2,v2}
\fmf{phantom,tension=2.8}{t3,v4}
\fmf{phantom,tension=1.05}{v3,b1}
\fmf{wiggly,left=.3}{v2,v4}
\fmf{wiggly,left=.9}{v2,v4}
\fmfv{decor.shape=circle,decor.filled=30,decor.size=9}{v1}
\fmfv{decor.shape=circle,decor.filled=30,decor.size=9}{v5}
\end{fmfgraph*}}+
\hspace{0.6cm}
\parbox{20mm}{
\begin{fmfgraph*}(50,50)
\fmfleft{v1}
\fmfright{v5}
\fmftop{t2,t4,t3}
\fmfbottom{b2,b1,b3}
\fmf{plain,left=0.35, tension=0.75}{v1,vb}
\fmf{plain,left=0.2, tension=114.25}{vb,v2}
\fmf{plain,left=0.4, tension=1}{v2,v3}
\fmf{plain,left=1., tension=1}{v3,v5}
\fmf{plain,left=0.35, tension=.75}{v5,vc}
\fmf{plain,left=0.2, tension=114.25}{vc,v4}
\fmf{plain,left=0.4, tension=1}{v4,v3}
\fmf{plain,left=1., tension=1}{v3,v1}
\fmf{phantom,tension=3.}{t4,v2}
\fmf{phantom,tension=6}{t2,vb}
\fmf{phantom,tension=3.}{b1,v4}
\fmf{phantom,tension=6}{b3,vc}
\fmf{wiggly,right=2.3,tension=1}{vc,v2}
\fmf{wiggly,left=2.3,tension=1}{v4,vb}
\fmfv{decor.shape=circle,decor.filled=30,decor.size=9}{v1}
\fmfv{decor.shape=circle,decor.filled=30,decor.size=9}{v5}
\end{fmfgraph*}}~~~~+
\hspace{0.6cm}
\parbox{20mm}{
\begin{fmfgraph*}(50,50)
\fmfleft{v1}
\fmfright{v5}
\fmftop{t2,t4,t3}
\fmfbottom{b2,b1,b3}
\fmf{plain,left=0.35, tension=0.75}{v1,vb}
\fmf{plain,left=0.2, tension=114.25}{vb,v2}
\fmf{plain,left=0.4, tension=1}{v2,v3}
\fmf{plain,left=1., tension=1}{v3,v5}
\fmf{plain,left=0.35, tension=.75}{v5,vc}
\fmf{plain,left=0.2, tension=114.25}{vc,v4}
\fmf{plain,left=0.4, tension=1}{v4,v3}
\fmf{plain,left=1., tension=1}{v3,v1}
\fmf{phantom,tension=3.}{t4,v2}
\fmf{phantom,tension=6}{t2,vb}
\fmf{phantom,tension=3.}{b1,v4}
\fmf{phantom,tension=6}{b3,vc}
\fmf{dashes,right=2.3,tension=1}{vc,v2}
\fmf{dashes,left=2.3,tension=1}{v4,vb}
\fmfv{decor.shape=circle,decor.filled=30,decor.size=9}{v1}
\fmfv{decor.shape=circle,decor.filled=30,decor.size=9}{v5}
\end{fmfgraph*}}~~~~+
\hspace{0.3cm}
\parbox{20mm}{
\begin{fmfgraph*}(50,50)
\fmfleft{v1}
\fmfright{v3}
\fmftop{t1,t2}
\fmfbottom{b1,b2}
\fmf{plain,left=.8, tension=1}{v1,v4,v5,v3,v4,v1}
\fmf{plain}{v4,v5}
\fmf{plain,right=.93}{v4,v5}
\fmf{phantom,tension=1}{t1,v4}
\fmf{phantom,tension=4}{t2,v5}
\fmfv{decor.shape=circle,decor.filled=30,decor.size=9}{v1}
\fmfv{decor.shape=circle,decor.filled=30,decor.size=9}{v3}
\end{fmfgraph*}}={\cal O}(\e).\]
\end{center}\vspace{0.9cm}
The remaining diagrams are individually finite and sum to 
\be
\text{sum of diagrams in Table \ref{tab:maindiags}} \stackrel{500}{\approx}
-\frac{1}{\sqrt{2}}\frac{(M+N)}{k^2}\frac{\pi^2}{192 (p^2)^{7-4\o}} ,
\ee
where the result has been verified to 500 digits of accuracy. Using
(\ref{unrenorm}) we find
\be
\hat C_{123}(\l,\hat\l) = 1-\l\hat\l\frac{\pi^2}{3} +{\cal O}(\l^2).
\ee
The two-point function evaluates to 
\be
\text{sum of diagrams in Table \ref{tab:twoptdiags}} \stackrel{500}{\approx}
(\l+\hat\l)^2\frac{1}{768 (p^2)^{7-5\o}} ,
\ee
and using (\ref{twoptnorm}) we find
\be
g_2(\l,\hat\l) = 1-\frac{\pi^2}{3}(\l+\hat\l)^2.
\ee
Finally using (\ref{renorm}) we obtain (\ref{mainresult}).

\section*{Acknowledgements}

I am first and foremost greatly indebted to Prof. Roman N. Lee not
only for his published basis of integrals and his reduction software
LiteRed \cite{Lee:2013mka}, but also for graciously providing me with
an expansion about three-dimensions for the master integrals out to
500 digit accuracy. Without these marvellous tools this project would
not have been possible. I would also like to acknowledge Claude Duhr,
Erik Panzer, Gordon Semenoff, and Christoph Sieg for discussions and
for helping me to navigate the literature on propagator integrals. I
have been supported by the consolidated grant ST/L000415/1
``String Theory, Gauge Theory \& Duality'' from the STFC.

\appendix

\section{Master integrals}
\label{app:4lmasters}

\begin{figure}
\begin{center}
\includegraphics[bb=76 286 519 628, clip=true]{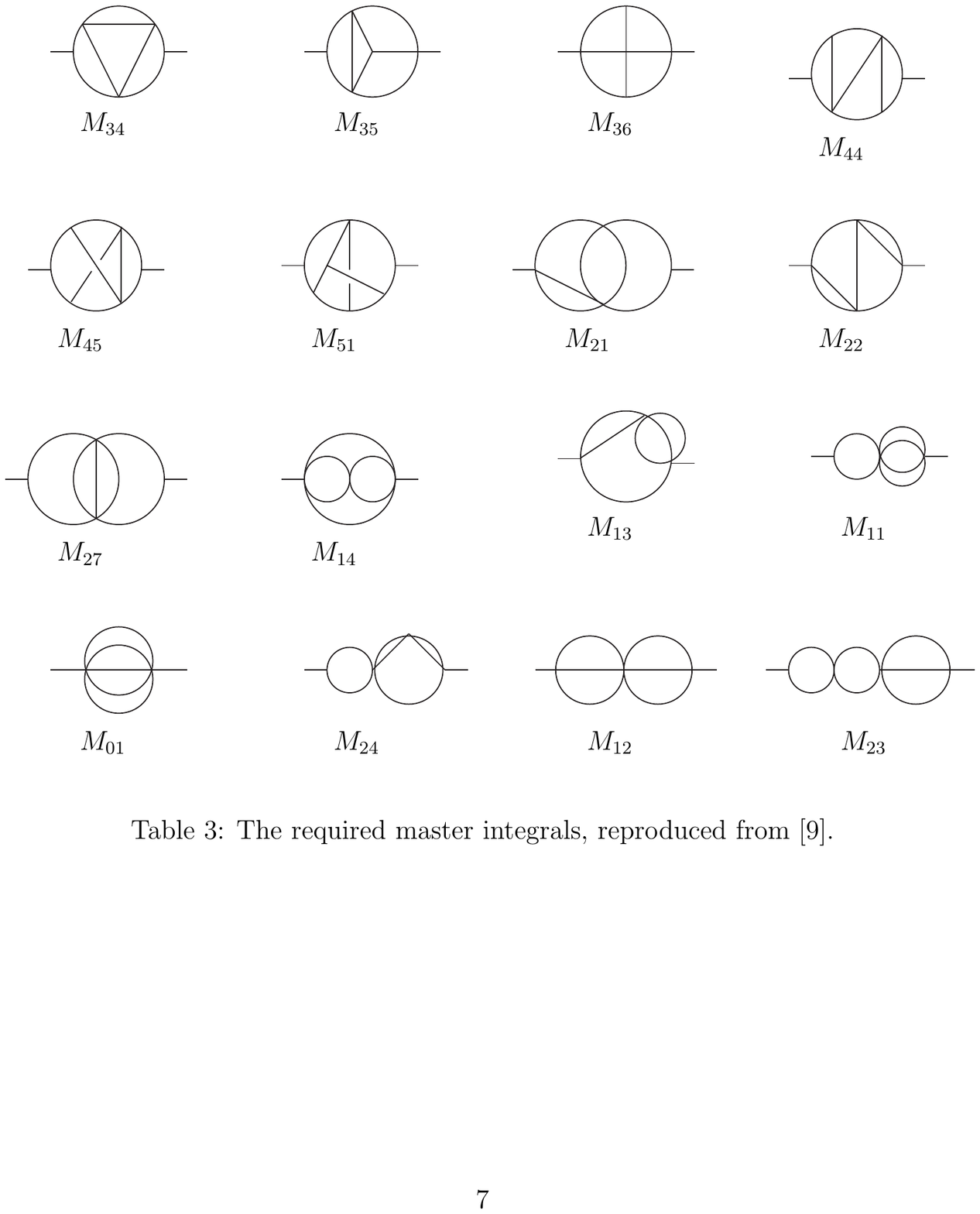}
\end{center}
\caption{The four-loop master integrals required for the three-point
  function, reproduced from \cite{Baikov:2010hf}.}
\label{fig:masters}
\end{figure}

The master integrals required for the calculation are a subset of the
basis presented in \cite{Baikov:2010hf}; they have been collected in
figure \ref{fig:masters}. The integrals $M_{21}$, $M_{22}$, $M_{34}$,
$M_{35}$, $M_{36}$, $M_{44}$, $M_{45}$, $M_{51}$ cannot be reduced to
$G$-functions or $F$-functions. Their values have been computed using
the dimensional recurrence analyticity method of Lee \cite{Lee:2009dh}
and their values out to 500 digits have been shared with me
\cite{LeePriv}. The algorithm PSLQ \cite{pslq} has been used in certain
places to provide analytical results for terms in the $\e$-expansion.

We use $d^{2\o}l/(2\pi)^{2\o}$ as the loop integration
measure where $d=2\o=3-2\e$. The $G$ and $F$ functions are given by 
\be
G(\a,\b) = \frac{1}{(4\pi)^\o}
\frac{\G(\a+\b-\o)\,\G(\o-\a)\,\G(\o-\b)}
{\G(\a)\,\G(\b)\,\G(2\o-\a-\b)},
\ee
and \cite{Kotikov:2003tc,Grozin:2003ak}
\bsp
&F_{\l} = \frac{2}{(4\pi)^{2\o}}\,\G(\o-1)\,\G(\o-\l-1)\,\G(\l-2\o+3)\Biggl(
-\frac{\pi\cot\left(\pi(2\o-\l)\right)}{\G(2\o-2)}\\
&+\frac{\G(\o-1)\,{}_3F_2\left(1,2+\l-\o,2\o-2;\l+1,\l-\o+3;1\right)}
{(\o-\l-2)\,\G(1+\l)\,\G(3\o-\l-4)}\,
\Biggr).
\end{split}
\ee
The Fourier transform is defined as 
\begin{equation}
\int \frac{d^{2\o}p}{(2\pi)^{2\o} } \frac{ e^{ip\cdot x} }{ [p^2]^s }
=\frac{\Gamma(\o-s)}{4^s\pi^\o\Gamma(s) }\frac{1}{[x^2]^{\o-s}}.
\end{equation} 
Below we employ a convenient normalization for the presentation of the
$\e$-expansion of the integrals
\be
{\cal M} = (4\pi)^{2-4\o}e^{-4\g\e}.
\ee
\bsp
&M_{34} = {\cal M} \left(-\frac{\pi^2}{32\e} +
\frac{\pi^2}{16}\left(3-2\log 2\right) + \left(\frac{37 \pi
  ^4}{96}-\frac{\pi^2}{8}  \bigl(2\log 2 (\log 2-3)+9\bigr)\right)\e
+{\cal O}(\e^2)\right),\\
&M_{35} = {\cal M} \Bigl( \frac{1}{16\e^2}+\frac{1}{16}(3\pi^2 - 60)
-10.1695380240468262091170104661503731482787\e\\
&+5.761640224581950185485181492971261298252179\e^2+{\cal
  O}(\e^3)\Bigr),\\
&M_{36}= {\cal M} \Bigl( \frac{1}{8\e^2} +\frac{1}{4\e}
+\frac{\pi^2}{24}-4
-31.24683122358910241344375025381651547680\e\\
&-104.97997949576156077918973047903786740\e^2
+{\cal O}(\e^3)
\Bigr),\\
&M_{44}= {\cal M} \Bigl( \frac{1}{8\e^2} +\frac{1}{4\e}
+\frac{5\pi^2}{24}-4
-8.154600887744206655212139477314414443076\e\\
&+65.5962689793108431307288614828060453748\e^2
+{\cal O}(\e^3)
\Bigr),\\
&M_{45}= {\cal M} \Bigl( \frac{1}{8\e^2}
+\frac{25\pi^2}{24}-\frac{31}{2}
-25.73103829909166269503717808059699905229\e\\
&+52.4026571263842398765902744496356247875\e^2
+{\cal O}(\e^3)
\Bigr),\\
&M_{51}= {\cal M} \Bigl( \frac{1}{8\e^2}
+\left(\frac{263}{120}-\frac{4\pi^2}{45}\right)\frac{1}{\e}
+1.3455583077571687466389820840805296520163\\
&-54.1517695706215470260568776499507551760225\e
+{\cal O}(\e^2)
\Bigr),\\
&M_{21} = {\cal M}\bigl(
5.462481216887406375320255581095582681248891250+{\cal
  O}(\e)\bigr),\\
&M_{22} = {\cal M}\bigl(
5.462481216887406375320255581095582681248891250+{\cal
  O}(\e)\bigr).\\
\end{split}
\ee
\bsp
&M_{27} = F_{3-2\o},\quad
M_{14} = G(1,1)^2\,G(1,4-2\o)\,G(1,5-3\o),\\
&M_{13} = G(1,1)^2\,G(1,2-\o)\,G(1,5-3\o),\quad
M_{11} = G(1,1)^2\,G(1,2-\o)\,G(1,3-2\o),\\
&M_{01} = G(1,1)\,G(1,2-\o)\,G(1,3-2\o)\,G(1,4-3\o),\\
&M_{24} = G(1,1)^3\,G(1,4-2\o),\quad
M_{12} = G^2(1,1)\,G^2(1,2-\o),\\
&M_{23} = G(1,1)^3\, G(1,2-\o).
\end{split}
\ee

\section{Reduction to master integrals}
\label{app:red}

\subsection{Three-point function diagrams}

I give below the reduction of the integrated three-point function in
terms of basis integrals. This reduction was performed using the code
LiteRed \cite{Lee:2013mka}. An overall factor of $(4\pi/k)^2$ is
suppressed. The colour factors of the diagrams are all equal to $(M+N)/\sqrt{2}$.

\vspace{0.85cm}
\[
\parbox{20mm}{
\begin{fmfgraph*}(50,50)
\fmfleft{v1}
\fmfright{v5}
\fmftop{t2,t4,t3}
\fmfbottom{b1}
\fmf{plain,left=0.5, tension=0.75}{v1,v2}
\fmf{plain,left=0.4, tension=1}{v2,v3}
\fmf{plain,left=0.4, tension=1}{v3,v4}
\fmf{plain,left=0.5, tension=0.75}{v4,v5}
\fmf{plain,left=1., tension=1}{v5,v3,v1}
\fmf{phantom,tension=2.8}{t2,v2}
\fmf{phantom,tension=2.8}{t3,v4}
\fmf{phantom,tension=1.05}{v3,b1}
\fmf{wiggly,left=.3}{v2,v4}
\fmf{wiggly,left=.9}{v2,v4}
\fmfv{decor.shape=circle,decor.filled=30,decor.size=9}{v1}
\fmfv{decor.shape=circle,decor.filled=30,decor.size=9}{v5}
\end{fmfgraph*}}=
\hspace{0.6cm}
\parbox{20mm}{
\begin{fmfgraph*}(50,50)
\fmfleft{v1}
\fmfright{v5}
\fmftop{t2,t4,t3}
\fmfbottom{b2,b1,b3}
\fmf{plain,left=0.35, tension=0.75}{v1,vb}
\fmf{plain,left=0.2, tension=114.25}{vb,v2}
\fmf{plain,left=0.4, tension=1}{v2,v3}
\fmf{plain,left=1., tension=1}{v3,v5}
\fmf{plain,left=0.35, tension=.75}{v5,vc}
\fmf{plain,left=0.2, tension=114.25}{vc,v4}
\fmf{plain,left=0.4, tension=1}{v4,v3}
\fmf{plain,left=1., tension=1}{v3,v1}
\fmf{phantom,tension=3.}{t4,v2}
\fmf{phantom,tension=6}{t2,vb}
\fmf{phantom,tension=3.}{b1,v4}
\fmf{phantom,tension=6}{b3,vc}
\fmf{wiggly,right=2.3,tension=1}{vc,v2}
\fmf{wiggly,left=2.3,tension=1}{v4,vb}
\fmfv{decor.shape=circle,decor.filled=30,decor.size=9}{v1}
\fmfv{decor.shape=circle,decor.filled=30,decor.size=9}{v5}
\end{fmfgraph*}}~~~~=-4
\hspace{0.6cm}
\parbox{20mm}{
\begin{fmfgraph*}(50,50)
\fmfleft{v1}
\fmfright{v5}
\fmftop{t2,t4,t3}
\fmfbottom{b2,b1,b3}
\fmf{plain,left=0.35, tension=0.75}{v1,vb}
\fmf{plain,left=0.2, tension=114.25}{vb,v2}
\fmf{plain,left=0.4, tension=1}{v2,v3}
\fmf{plain,left=1., tension=1}{v3,v5}
\fmf{plain,left=0.35, tension=.75}{v5,vc}
\fmf{plain,left=0.2, tension=114.25}{vc,v4}
\fmf{plain,left=0.4, tension=1}{v4,v3}
\fmf{plain,left=1., tension=1}{v3,v1}
\fmf{phantom,tension=3.}{t4,v2}
\fmf{phantom,tension=6}{t2,vb}
\fmf{phantom,tension=3.}{b1,v4}
\fmf{phantom,tension=6}{b3,vc}
\fmf{dashes,right=2.3,tension=1}{vc,v2}
\fmf{dashes,left=2.3,tension=1}{v4,vb}
\fmfv{decor.shape=circle,decor.filled=30,decor.size=9}{v1}
\fmfv{decor.shape=circle,decor.filled=30,decor.size=9}{v5}
\end{fmfgraph*}}~~~
=M_{11} \left(\frac{1}{2 \o-3}+2\right)\]
\vspace{0.65cm}
\[+M_{14} \left(\frac{5}{12 (3
  \o-4)}-\frac{25}{12}\right)+M_{34} \left(\frac{1}{4}-\frac{1}{4 (3
  \o-4)}\right),
\]
\vspace{0.75cm}
\[%
\parbox{20mm}{
\begin{fmfgraph*}(50,50)
\fmfleft{v1}
\fmfright{v5}
\fmftop{t2,t4,t3}
\fmfbottom{b1}
\fmf{plain,left=0.35, tension=0.75}{v1,vb}
\fmf{plain,left=0.1, tension=0.25}{vb,v2}
\fmf{plain,left=0.4, tension=1}{v2,v3}
\fmf{plain,left=0.4, tension=1}{v3,v4}
\fmf{plain,left=0.5, tension=0.75}{v4,v5}
\fmf{plain,left=1., tension=1}{v5,v3,v1}
\fmf{phantom,tension=2.8}{t2,v2}
\fmf{phantom,tension=1.}{t2,vb}
\fmf{phantom,tension=2.8}{t3,v4}
\fmf{phantom,tension=1.05}{v3,b1}
\fmf{wiggly,left=.3}{v2,v4}
\fmf{wiggly,left=.7,tension=0.1}{vb,v4}
\fmfv{decor.shape=circle,decor.filled=30,decor.size=9}{v1}
\fmfv{decor.shape=circle,decor.filled=30,decor.size=9}{v5}
\end{fmfgraph*}} =
\hspace{0.5cm}
\parbox{20mm}{
\begin{fmfgraph*}(50,50)
\fmfleft{v1}
\fmfright{v5}
\fmftop{t2,t4,t3}
\fmfbottom{b2,b1,b3}
\fmf{plain,left=0.35, tension=0.75}{v1,vb}
\fmf{plain,left=0.2, tension=114.25}{vb,v2}
\fmf{plain,left=0.4, tension=1}{v2,v3}
\fmf{plain,left=1., tension=1}{v3,v5}
\fmf{plain,left=0.35, tension=.75}{v5,vc}
\fmf{plain,left=0.2, tension=4.25}{vc,v4}
\fmf{plain,left=0.3, tension=1}{v4,v3}
\fmf{plain,left=1., tension=1}{v3,v1}
\fmf{phantom,tension=3.}{t4,v2}
\fmf{phantom,tension=6}{t2,vb}
\fmf{phantom,tension=3.}{b1,v4}
\fmf{phantom,tension=6}{b3,vc}
\fmf{wiggly,right=1.93,tension=1}{vc,v2}
\fmf{wiggly,left=1.93,tension=1}{v4,vb}
\fmfv{decor.shape=circle,decor.filled=30,decor.size=9}{v1}
\fmfv{decor.shape=circle,decor.filled=30,decor.size=9}{v5}
\end{fmfgraph*}} ~~~
=M_{11} \left(-\frac{1}{4 (2 \o-3)^2}-\frac{15}{4 (2
  \o-3)}-\frac{13}{2}\right)\]
\vspace{0.65cm}
\[+M_{14} \left(-\frac{5}{4 (3
  \o-4)}+\frac{1}{2 (2 \o-3)}+\frac{35}{4}\right)
+M_{34} \left(\frac{3}{4 (3 \o-4)}-\frac{3}{4}\right)+M_{27},\]

\[\parbox{20mm}{
\begin{fmfgraph*}(50,50)
\fmfleft{v1}
\fmfright{v5}
\fmftop{t2,t4,t3}
\fmfbottom{b2,b1,b3}
\fmf{plain,left=0.5, tension=0.75}{v1,v2}
\fmf{plain,left=0.25, tension=2}{v2,v3}
\fmf{plain,left=0.4, tension=1}{v3,v4}
\fmf{plain,left=0.4, tension=0.75}{v4,v5}
\fmf{plain,left=1.1, tension=1}{v5,v3}
\fmf{plain,left=0.6, tension=0.75}{v3,v6}
\fmf{plain,left=0.4, tension=1}{v6,v1}
\fmf{phantom,tension=2.2}{t2,v2}
\fmf{phantom,tension=4.2}{t3,v4}
\fmf{phantom,tension=.35}{v3,b1}
\fmf{phantom,tension=6.8}{v6,b2}
\fmf{wiggly,left=.8}{v2,v4}
\fmf{wiggly,left=2.05}{v6,v4}
\fmfv{decor.shape=circle,decor.filled=30,decor.size=9}{v1}
\fmfv{decor.shape=circle,decor.filled=30,decor.size=9}{v5}
\end{fmfgraph*}} = 
M_{01} \left(\frac{2}{(2 \o-3)^2}-\frac{6752}{27 (3 \o-5)}-\frac{2240}{27
  (3 \o-5)^2}+\frac{560}{3 (5 \o-8)}-\frac{2}{3 (2
  \o-3)}-\frac{1340}{27}\right)\]\[+M_{11} \left(\frac{13}{6 (2
  \o-3)^2}+\frac{20}{3 \o-5}+\frac{112}{15 (5 \o-8)}+\frac{7}{2 (2
  \o-3)}+\frac{93}{5}\right)\]\[+M_{13} \left(-\frac{1}{3 (2
  \o-3)^2}-\frac{16}{3 (3 \o-5)}-\frac{16}{3 (5 \o-8)}-\frac{4}{3 (2
  \o-3)}-\frac{31}{3}\right)\]\[+M_{14} \left(-\frac{100}{27 (3
  \o-5)}-\frac{16}{27 (3 \o-5)^2}-\frac{1}{2
  \o-3}-\frac{235}{27}\right)\]\[+2 M_{22} \left(-\frac{8}{15 (5
  \o-8)}-\frac{8}{9 (3 \o-5)}+\frac{52}{45}\right)+M_{27}
\left(-\frac{4}{3 (3 \o-5)}-\frac{4}{3 (2
  \o-3)}-\frac{10}{3}\right)\]\[+M_{35} \left(\frac{1}{3 (2
  \o-3)^2}+\frac{4}{15 (5 \o-8)}-\frac{4}{3 (2
  \o-3)}+\frac{13}{15}\right)+M_{45} \left(-\frac{1}{12 (2
  \o-3)^2}+\frac{1}{6 (2 \o-3)}-\frac{1}{12}\right),\]

\[\parbox{20mm}{
\begin{fmfgraph*}(50,50)
\fmfleft{v1}
\fmfright{v5}
\fmftop{t2,t4,t3}
\fmfbottom{b2,b1,b3}
\fmf{plain,left=0.35, tension=0.75}{v1,vb}
\fmf{plain,left=0.2, tension=4.25}{vb,v2}
\fmf{plain,left=0.3, tension=1}{v2,v3}
\fmf{plain,left=1., tension=1}{v3,v5}
\fmf{plain,left=0.35, tension=.75}{v5,vc}
\fmf{plain,left=0.2, tension=4.25}{vc,v4}
\fmf{plain,left=0.3, tension=1}{v4,v3}
\fmf{plain,left=1., tension=1}{v3,v1}
\fmf{phantom,tension=3.}{t4,v2}
\fmf{phantom,tension=6}{t2,vb}
\fmf{phantom,tension=3.}{b1,v4}
\fmf{phantom,tension=6}{b3,vc}
\fmf{wiggly,right=1.93,tension=1}{vc,v2}
\fmf{wiggly,left=1.93,tension=1}{v4,vb}
\fmfv{decor.shape=circle,decor.filled=30,decor.size=9}{v1}
\fmfv{decor.shape=circle,decor.filled=30,decor.size=9}{v5}
\end{fmfgraph*}} ~~~=
M_{01} \Bigl(\frac{19691}{16 (\o-2)^2}+\frac{1575}{8 (\o-2)^3}+\frac{45}{2
  (\o-2)^4}+\frac{2333}{32 (2 \o-3)}-\frac{17}{(2
  \o-3)^2}\]\[-\frac{3}{(2 \o-3)^3}+\frac{280}{3 (3
  \o-5)}+\frac{55531}{16 (4 \o-7)}+\frac{1617}{4 (4 \o-7)^2}-\frac{560}{5
  \o-8}+\frac{10711}{16 (\o-2)}+\frac{65185}{96}\Bigr)\]\[+ M_{11}
\left(\frac{29}{(\o-2)^2}+\frac{6}{(\o-2)^3}+\frac{2}{2 \o-3}-\frac{2}{(2
  \o-3)^2}-\frac{10}{4 \o-7}-\frac{112}{5 (5
  \o-8)}+\frac{52}{\o-2}+\frac{131}{5}\right)\]\[ M_{12}
\Bigl(\frac{14}{(\o-2)^2}+\frac{7}{2 (\o-2)^3}-\frac{1}{2 (2
  \o-3)}-\frac{1}{8 (2 \o-3)^2}-\frac{1155}{64 (4 \o-7)}-\frac{175}{128
  (4 \o-7)^2}\]\[-\frac{8}{5 (5 \o-8)}+\frac{231}{8
  (\o-2)}+\frac{7497}{640}\Bigr)+M_{13} \Bigl(-\frac{163}{4
  (\o-2)^2}-\frac{33}{4 (\o-2)^3}+\frac{4}{2 \o-3}+\frac{3}{4 (2
  \o-3)^2}\]\[+\frac{3285}{128 (4 \o-7)}+\frac{525}{256 (4
  \o-7)^2}+\frac{16}{5 \o-8}-\frac{1323}{16
  (\o-2)}-\frac{7239}{256}\Bigr)\]\[ +M_{14} \left(\frac{8}{3 (3
  \o-5)}+\frac{10}{3 (3 \o-4)}+\frac{115}{4 (4 \o-7)}+\frac{105}{16 (4
  \o-7)^2}-\frac{15}{\o-2}-\frac{1015}{48}\right)\]\[ +2 M_{22}
\left(-\frac{2}{(\o-2)^2}-\frac{2}{4 \o-7}+\frac{8}{5 (5
  \o-8)}-\frac{6}{\o-2}-\frac{34}{5}\right)\]\[+M_{27}
\left(-\frac{7}{(\o-2)^2}+\frac{3}{2 \o-3}-\frac{35}{4 (4
  \o-7)}-\frac{35}{2 (\o-2)}-\frac{43}{4}\right)\]\[+M_{34}
\left(\frac{1}{2 (2 \o-3)}-\frac{2}{3 \o-4}+\frac{5}{16 (4
  \o-7)}-\frac{5}{4 (\o-2)}-\frac{37}{16}\right)\]\[+M_{35}
\left(\frac{1}{2 (2 \o-3)^2}+\frac{3}{16 (4 \o-7)}-\frac{4}{5 (5
  \o-8)}-\frac{7}{8 (2 \o-3)}+\frac{87}{80}\right)\]\[+M_{36}
\left(-\frac{1}{8 (2 \o-3)^2}-\frac{5}{128 (4 \o-7)}+\frac{13}{512 (4
  \o-7)^2}+\frac{2}{5 (5 \o-8)}-\frac{23}{128 (2
  \o-3)}+\frac{303}{2560}\right)\]\[+M_{51} \left(-\frac{3}{128 (4
  \o-7)}-\frac{21}{512 (4 \o-7)^2}+\frac{15}{128 (2
  \o-3)}-\frac{27}{512}\right),\]

\[\parbox{20mm}{
\begin{fmfgraph*}(50,50)
\fmfleft{v1}
\fmfright{v3}
\fmf{plain,left=0.3, tension=1}{v1,v4,v3,v2,v1}
\fmffixed{(0,33)}{v2,v4}
\fmf{plain,left=.93}{v2,v5}
\fmf{dashes,left=.93}{v5,v4}
\fmf{plain,right=.93}{v2,v5}
\fmf{dashes,right=.93}{v5,v4}
\fmfv{decor.shape=circle,decor.filled=30,decor.size=9}{v1}
\fmfv{decor.shape=circle,decor.filled=30,decor.size=9}{v3}
\end{fmfgraph*}}= -2\,G(1,1)\,F_{3-2\o},\]

\[\parbox{20mm}{
\begin{fmfgraph*}(50,50)
\fmfleft{v1}
\fmfright{v3}
\fmftop{t1,t2}
\fmfbottom{b1,b2}
\fmf{plain,left=.8, tension=1}{v1,v4,v5,v3,v4,v1}
\fmf{plain}{v4,v5}
\fmf{plain,right=.93}{v4,v5}
\fmf{phantom,tension=1}{t1,v4}
\fmf{phantom,tension=4}{t2,v5}
\fmfv{decor.shape=circle,decor.filled=30,decor.size=9}{v1}
\fmfv{decor.shape=circle,decor.filled=30,decor.size=9}{v3}
\end{fmfgraph*}} =  G^2(1, 1)\,G(1, 2 - \o)\,G(1, 4 - 2\o).\]

\subsection{Two-point function diagrams}

The two-point function of the length-four operator is essential for
renormalizing the three-point function. Here I give the reduction to
master integrals which was performed in a very similar context in
\cite{Young:2014lka}. Again, a factor of $(4\pi/k)^2$ is suppressed
while the colour factors for the various diagrams are
\be
T_1,T_2,T_4,T_6 = 2(M^2+N^2),\qquad
T_3,T_5,T_8,T_9 = 4MN.
\ee
The scalar propagator at two loops involves the quantity \cite{Minahan:2009wg}
\bsp
Z_\text{scalar}& = -\frac{1}{(4\pi)^2} \Biggl[
MN\left(\frac{3}{4(3-2\o)}+\frac{1}{4}\left(
-\frac{3\pi^2}{2} +25 -3\g +3\log(4\pi) \right)\right)\\
&+\frac{(M-N)^2}{4}\left(\frac{1}{2(3-2\o)}
-\frac{\pi^2}{4}+\frac{1}{2}\left(
3-\g+\log(4\pi)\right)\right)\Biggr]+{\cal O}(\e).
\end{split}
\ee
The following three-loop master integrals are employed here 
\vspace{-0.5cm}
\[P_1 = 
\parbox{20mm}{
\begin{fmfgraph*}(70,70)
\fmfleft{i}
\fmfright{o}
\fmf{plain,tension=4}{i,v1}
\fmf{plain,tension=4}{v3,o}
\fmf{plain,left=1}{v1,va,v3,va,v1}
\fmffreeze
\fmf{plain}{v1,va}
\end{fmfgraph*}}
\quad
=G^2(1,1)\,G(1,2-\o),\quad
P_2=
\parbox{20mm}{
\begin{fmfgraph*}(50,50)
\fmfleft{i}
\fmfright{o}
\fmf{plain,tension=4}{i,v1}
\fmf{plain,tension=4}{v3,o}
\fmf{plain,left=1}{v1,va,v3,va,v1}
\fmffreeze
\fmf{plain,left=1.}{v1,v3}
\end{fmfgraph*}}
=G^2(1,1)\,G(1,4-2\o),\]
\vspace{-1cm}
\[P_3=
\parbox{20mm}{
\begin{fmfgraph*}(60,60)
\fmfleft{i}
\fmfright{o}
\fmftop{t}
\fmfbottom{b}
\fmf{phantom,tension=1.7}{t,va}
\fmf{phantom,tension=1.7}{b,vb}
\fmf{plain,tension=4}{i,v1}
\fmf{plain,tension=4}{v3,o}
\fmf{plain,left=0.3}{v1,va,v3,vb,v1}
\fmffreeze
\fmf{plain,right=0.5}{va,vb,va}
\end{fmfgraph*}}
\,=G(1,1)\,F_{2-\o},~
P_4=
\parbox{20mm}{
\begin{fmfgraph*}(60,60)
\fmfleft{i}
\fmfright{o}
\fmf{plain,tension=8}{i,v1}
\fmf{plain,tension=8}{v3,o}
\fmf{plain,left=1}{v1,v3,v1}
\fmf{plain,left=0.5}{v1,v3,v1}
\end{fmfgraph*}}
\,=G(1,1)\,G(1,2-\o)\,G(1,3-2\o),\]
\vspace{-0.5cm}
\[
\hspace{-0.5cm}
P_6=
\parbox{20mm}{
\begin{fmfgraph*}(60,60)
\fmfleft{i}
\fmfright{o}
\fmf{plain,tension=4}{i,v1}
\fmf{plain,tension=4}{v3,o}
\fmf{plain,left=1}{v1,v3}
\fmf{plain,tension=2}{v1,va}
\fmf{plain}{va,vb}
\fmf{plain,tension=2}{vb,v3}
\fmffreeze
\fmf{plain,left=1.}{va,vb,va}
\end{fmfgraph*}}\,=
G(1,1)\,G(1,2-\o)\,G(1,5-2\o),
\hspace{0.3cm}
P_7=
\parbox{20mm}{
\begin{fmfgraph*}(50,50)
\fmfleft{i}
\fmfright{o}
\fmf{plain,tension=4}{i,v1}
\fmf{plain,tension=4}{v3,o}
\fmf{plain,tension=3}{v1,vc}
\fmf{plain,tension=3}{vd,v3}
\fmf{plain,left=1}{vc,va,vd,va,vc}
\fmffreeze
\fmf{plain,left=.9}{v1,v3}
\end{fmfgraph*}}\,=G^2(1,1)\,G(1,6-2\o).\]
\vspace{-0.5cm}
\[
\hspace{-5.95cm}
\]
The reductions are as follows
\[T_1=~~\parbox{20mm}{
\begin{fmfgraph*}(50,50)
\fmfleft{v1}
\fmfright{v3}
\fmf{plain,left=0.3, tension=1}{v1,v4,v3,v2,v1}
\fmf{plain,left=1., tension=1}{v1,v3,v1}
\fmffixed{(0,33)}{v2,v4}
\fmf{wiggly,left=.3}{v2,v4,v2}
\fmfv{decor.shape=circle,decor.filled=30,decor.size=9}{v1}
\fmfv{decor.shape=circle,decor.filled=30,decor.size=9}{v3}
\end{fmfgraph*}} = G(1,5-3\o)\,G(1,6-4\o)\left(\frac{P_3 (3-2 \omega )}{4 (3 \omega -4)}
+\frac{1}{6} P_4
\left(\frac{1}{3 \omega -4}+\frac{3}{2 \omega -3}+10\right)\right),\]

\[T_2=~~\parbox{20mm}{
\begin{fmfgraph*}(50,50)
\fmfleft{v1}
\fmftop{t}
\fmfright{v2}
\fmfbottom{b1,b2}
\fmf{phantom,tension=2.95}{t,va}
\fmf{phantom,tension=2.15}{b1,vb1}
\fmf{phantom,tension=2.15}{b2,vb2}
\fmf{plain,left=0.3}{v1,va}
\fmf{plain,left=0.3}{va,v2}
\fmf{plain,right=0.25}{v1,vb1}
\fmf{plain,right=0.25}{vb1,vb2}
\fmf{plain,right=0.25}{vb2,v2}
\fmffreeze
\fmf{wiggly}{va,vb1}
\fmf{wiggly}{va,vb2}
\fmf{plain,left=1., tension=1}{v1,v2}
\fmf{plain,left=1.2, tension=1}{v2,v1}
\fmfv{decor.shape=circle,decor.filled=30,decor.size=9}{v1}
\fmfv{decor.shape=circle,decor.filled=30,decor.size=9}{v2}
\end{fmfgraph*}} 
= G(1,5-3\o)\,G(1,6-4\o)
\Biggl(\frac{P_3 (9-6 \omega )}{16-12 \omega }+P_4 \left(\frac{1}{8-6 \omega
}+\frac{1}{3-2 \omega }-4\right)\]\[+P_1-\frac{P_2}{2}\Biggr),
\]

\[T_3=~~~\parbox{20mm}{
\begin{fmfgraph*}(50,50)
\fmfleft{v1}
\fmftop{t1,t2,t3}
\fmfright{v2}
\fmfbottom{b}
\fmf{phantom,tension=1.0}{t1,va1}
\fmf{phantom,tension=1.1}{t2,va2}
\fmf{phantom,tension=1.0}{t3,va3}
\fmf{phantom,tension=2.95}{b,vb}
\fmf{plain,left=0.17}{v1,va1}
\fmf{plain,left=0.15}{va1,va2}
\fmf{plain,left=0.15}{va2,va3}
\fmf{plain,left=0.17}{va3,v2}
\fmf{plain,right=0.3}{v1,vb}
\fmf{plain,right=0.3}{vb,v2}
\fmffreeze
\fmf{wiggly,left=0.75}{va1,va3}
\fmf{wiggly}{va2,vb}
\fmf{plain,left=1.3, tension=1}{v1,v2}
\fmf{plain,left=1., tension=1}{v2,v1}
\fmfv{decor.shape=circle,decor.filled=30,decor.size=9}{v1}
\fmfv{decor.shape=circle,decor.filled=30,decor.size=9}{v2}
\end{fmfgraph*}} = 2T_4=
\hspace{0.4cm} 
\parbox{20mm}{
\begin{fmfgraph*}(50,50)
\fmfleft{v1}
\fmftop{t1,t2,t3}
\fmfright{v2}
\fmfbottom{b}
\fmf{phantom,tension=1.0}{t1,va1}
\fmf{phantom,tension=1.1}{t2,va2}
\fmf{phantom,tension=1.0}{t3,va3}
\fmf{phantom,tension=2.95}{b,vb}
\fmf{plain,left=0.17}{v1,va1}
\fmf{plain,left=0.15}{va1,va2}
\fmf{plain,left=0.15}{va2,va3}
\fmf{plain,left=0.17}{va3,v2}
\fmf{plain,right=0.3}{v1,vb}
\fmf{plain,right=0.3}{vb,v2}
\fmffreeze
\fmf{wiggly}{va1,vcen}
\fmf{wiggly}{va3,vcen}
\fmf{wiggly}{vb,vcen}
\fmf{plain,left=1.2, tension=1}{v1,v2}
\fmf{plain,left=1., tension=1}{v2,v1}
\fmfv{decor.shape=circle,decor.filled=30,decor.size=9}{v1}
\fmfv{decor.shape=circle,decor.filled=30,decor.size=9}{v2}
\end{fmfgraph*}}= G(1,5-3\o)\,G(1,6-4\o)\Biggl(
P_1 \left(\frac{2}{2 \omega -3}+4\right)\]\[+\frac{P_3 (5-4 \omega )}{4-3
  \omega }
-\frac{2 P_4 (\omega -1) (4 \omega -5) (8 \omega -11)}{(3-2
  \omega )^2 (3 \omega -4)}-4 P_2,
\Biggr)\]

\[T_5=~~~\parbox{20mm}{
\begin{fmfgraph*}(50,50)
\fmfleft{v1}
\fmfright{v3}
\fmf{plain,left=0.3, tension=1}{v1,v4,v3,v2,v1}
\fmffixed{(0,33)}{v2,v4}
\fmf{wiggly}{v2,vc,v4}
\fmf{plain,left=1., tension=1}{v1,v3,v1}
\fmfv{d.sh=circle,l.d=0, d.f=empty,d.si=.25w,l=$1$}{vc}
\fmfv{decor.shape=circle,decor.filled=30,decor.size=9}{v1}
\fmfv{decor.shape=circle,decor.filled=30,decor.size=9}{v3}
\end{fmfgraph*}} = G(1,5-3\o)\,G(1,6-4\o)\Biggl(
\frac{P_3 (5-4 \omega )}{3 \omega -4}+P_4 \left(\frac{2}{12-9 \omega
}+\frac{12}{\omega -2}+\frac{40}{3}\right)
\Biggr),\]

\[T_6=~~~
\parbox{20mm}{
\begin{fmfgraph*}(50,50)
\fmfleft{v1}
\fmfright{v3}
\fmf{plain,left=1., tension=1}{v1,va}
\fmf{dashes,left=1., tension=1}{va,vb}
\fmf{plain,left=1., tension=1}{vb,v3}
\fmf{plain,right=1., tension=1}{v1,va}
\fmf{dashes,right=1., tension=1}{va,vb}
\fmf{plain,right=1., tension=1}{vb,v3}
\fmf{plain,left=.75, tension=1}{v1,v3,v1}
\fmfv{decor.shape=circle,decor.filled=30,decor.size=9}{v1}
\fmfv{decor.shape=circle,decor.filled=30,decor.size=9}{v3}
\end{fmfgraph*}} = -G^3(1,1)\,G(1,5-3\o)\,G(1,6-4\o),\]

\[T_7=~~~\parbox{20mm}{
\begin{fmfgraph*}(50,50)
\fmfleft{v1}
\fmfright{v3}
\fmf{plain,left=0.3, tension=1}{v1,v4,v3,v2,v1}
\fmffixed{(0,30)}{v2,v4}
\fmf{plain,left=1., tension=1}{v1,v3,v1}
\fmfv{d.sh=circle,l.d=0, d.f=empty,d.si=.25w,l=$2$}{v4}
\fmfv{decor.shape=circle,decor.filled=30,decor.size=9}{v1}
\fmfv{decor.shape=circle,decor.filled=30,decor.size=9}{v3}
\end{fmfgraph*}}=
4\,G(1,5-3\o)\,G(1,6-4\o) \,G(1,4-2\o)\,Z_\text{scalar},
\]

\[T_8=~~~\parbox{20mm}{
\begin{fmfgraph*}(50,50)
\fmfleft{v1}
\fmfright{v3}
\fmf{plain,left=0.93, tension=1}{v1,v4}
\fmf{plain, tension=1}{v1,v4}
\fmf{plain,right=0.93, tension=1}{v1,v4}
\fmf{plain,left=0.93, tension=1}{v4,v3}
\fmf{plain, tension=1}{v4,v3}
\fmf{plain,right=0.93, tension=1}{v4,v3}
\fmf{plain,left=1., tension=1}{v3,v1}
\fmfv{decor.shape=circle,decor.filled=30,decor.size=9}{v1}
\fmfv{decor.shape=circle,decor.filled=30,decor.size=9}{v3}
\end{fmfgraph*}} = 
\frac{1}{2}\,G^2(1, 1)\,G^2(1, 2 - \o)\,G(1,6-4\o) ,
\]

\[T_9=~~~\parbox{20mm}{
\begin{fmfgraph*}(50,50)
\fmfleft{v1}
\fmfright{v3}
\fmftop{t1}
\fmfbottom{b1}
\fmf{plain,left=0.3, tension=1}{v1,v4,v3}
\fmf{plain, tension=1}{v1,v2,v3}
\fmf{plain,right=.3, tension=1}{v1,v5,v3}
\fmf{plain,right=1, tension=1}{v1,v3}
\fmf{phantom,tension=5}{b1,v5}
\fmf{phantom,tension=5}{t1,v4}
\fmf{wiggly}{v2,v5}
\fmf{wiggly}{v2,v4}
\fmfv{decor.shape=circle,decor.filled=30,decor.size=9}{v1}
\fmfv{decor.shape=circle,decor.filled=30,decor.size=9}{v3}
\end{fmfgraph*}} = G(1,6-4\o) 
\Biggl(
M_{01} \left(\frac{9}{(2 \o-3)^2}+\frac{140}{3 (3 \o-5)}+\frac{2}{3 (3
  \o-4)}+\frac{18}{2 \o-3}+40\right)\]\[+M_{12} \left(-\frac{16}{25 (5
  \o-8)}-\frac{14}{25 (5 \o-7)}+\frac{3}{2 (2
  \o-3)}+\frac{46}{25}\right)+M_{13} \left(-\frac{4}{2
  \o-3}-10\right)\]\[+M_{14} \left(\frac{4}{3 (3
  \o-5)}+\frac{17}{3}\right)+M_{36} \left(\frac{4}{25 (5
  \o-8)}+\frac{36}{25 (5 \o-7)}-\frac{1}{2 (2 \o-3)}-\frac{3}{50}\right)
\Biggr).
\]

\end{fmffile}
\bibliography{extremal}%
\end{document}